\documentclass[aps,twocolumn,superscriptaddress]{revtex4-1}

\usepackage{epsfig}
\usepackage{microtype}

\usepackage{amssymb} 
\usepackage{graphicx} 
\usepackage{color}  
\usepackage[normalem]{ulem}
\usepackage{comment} 
\usepackage{units}
\usepackage{sidecap} 
\usepackage{textcomp} 

\usepackage{bigstrut}
\usepackage{placeins}
\usepackage{soul}
\usepackage[usenames,dvipsnames]{xcolor}

\usepackage[bookmarks=false,colorlinks]{hyperref}
\hypersetup{
	linkcolor=MidnightBlue,        
	citecolor=blue,     
	filecolor=Plum,      	
	urlcolor=MidnightBlue,           
}

\def \ybco{YBa$_2$Cu$_3$O$_{6+\delta}$}

\def \ybcosev{YBa$_2$Cu$_3$O$_{7}$}
\def \prybco{Pr$_x$Y$_{1-x}$Ba$_2$Cu$_3$O$_{6+\delta}$}
\def \prybcosev{Pr$_x$Y$_{1-x}$Ba$_2$Cu$_3$O$_{7}$}

\def \prbco{PrBa$_2$Cu$_3$O$_{6+\delta}$}
\def \dybcosix{DyBa$_2$Cu$_3$O$_{6}$}
\def \prbcosix{PrBa$_2$Cu$_3$O$_{6}$}

\definecolor{yilu}{rgb}{0.20, 0.30, 0.54}

\definecolor{bgunn}{rgb}{0.38, 0.55, 0.37}

\begin{document}
\title{Stabilization of three-dimensional charge order through interplanar orbital hybridization in \prybco}

\author{Alejandro Ruiz}
\email[These authors contributed equally to this work.]{}
\affiliation{Department of Physics, Center for Advanced Nanoscience, University of California, San Diego, California 92093, USA}
\affiliation{Massachusetts Institute of Technology, Cambridge, Massachusetts 02138, USA}

\author{Brandon Gunn}
\email[These authors contributed equally to this work.]{}
\affiliation{Department of Physics, Center for Advanced Nanoscience, University of California, San Diego, California 92093, USA}

\author{Yi Lu}
\affiliation{Institute for Theoretical Physics, Heidelberg University, Philosophenweg 19, 69120 Heidelberg, Germany}

\author{Kalyan Sasmal}
\affiliation{Department of Physics, Center for Advanced Nanoscience, University of California, San Diego, California 92093, USA}

\author{Camilla M. Moir}
\affiliation{Department of Physics, Center for Advanced Nanoscience, University of California, San Diego, California 92093, USA}

\author{Rourav Basak}
\affiliation{Department of Physics, Center for Advanced Nanoscience, University of California, San Diego, California 92093, USA}

\author{Hai Huang}
\email[Present address: Department of Materials Science, Fudan University, 220 Handan Road, Shanghai, 200433, China]{}
\affiliation{Stanford Synchrotron Radiation Lightsource, SLAC National Accelerator Laboratory, Menlo Park, California 94025, USA}

\author{Jun-Sik Lee}
\affiliation{Stanford Synchrotron Radiation Lightsource, SLAC National Accelerator Laboratory, Menlo Park, California 94025, USA}

\author{Fanny Rodolakis}
\affiliation{Advanced Photon Source, Argonne National Laboratory, Argonne, Illinois 60439, USA}

\author{Timothy\,J.\,Boyle}
\affiliation{Department of Physics, University of California, Davis, California 95616, USA}
\affiliation{Department of Physics, Yale University, New Haven, Connecticut 06520, USA}

\author{Morgan\,Walker}
\affiliation{Department of Physics, University of California, Davis, California 95616, USA}

\author{Yu He}
\affiliation{Department of Applied Physics, Yale University, New Haven, Connecticut 06511, USA}

\author{Santiago\,Blanco-Canosa}
\affiliation{Donostia International Physics Center, DIPC, 20018 Donostia-San Sebastian, Basque Country, Spain}
\affiliation{IKERBASQUE, Basque Foundation for Science, 48013 Bilbao, Spain}

\author{Eduardo\,H.\,da Silva Neto}
\affiliation{Department of Physics, University of California, Davis, California 95616, USA}
\affiliation{Department of Physics, Yale University, New Haven, Connecticut 06520, USA}
\affiliation{Energy Sciences Institute, Yale University, West Haven, Connecticut 06516, USA}

\author{M. Brian Maple}
\affiliation{Department of Physics, Center for Advanced Nanoscience, University of California, San Diego, California 92093, USA}

\author{Alex Frano}
\email[]{afrano@ucsd.edu}
\affiliation{Department of Physics, Center for Advanced Nanoscience, University of California, San Diego, California 92093, USA}

\date{\today}

\begin{abstract}
\textbf{The shape of 3$d$-orbitals often governs the electronic and magnetic properties of correlated transition metal oxides. In the superconducting cuprates, the planar confinement of the $d_{x^2-y^2}$ orbital dictates the two-dimensional nature of the unconventional superconductivity and a competing charge order. Achieving orbital-specific control of the electronic structure to allow coupling pathways across adjacent planes would enable direct assessment of the role of dimensionality in the intertwined orders. Using Cu-$L_3$ and Pr-$M_5$ resonant x-ray scattering and first principles calculations, we report a highly correlated three-dimensional charge order in Pr-substituted \ybcosev, where the Pr $f$-electrons create a direct orbital bridge between CuO$_2$ planes. With this we demonstrate that interplanar orbital engineering can be used to surgically control electronic phases in correlated oxides and other layered materials.}
\end{abstract}

\pacs{71.18.+y,74.72.-h,72.15.Gd}
\maketitle

The cuprate phase diagram illustrates a quintessential example of a low-dimensional correlated quantum system: a multitude of fascinating electronic phases, including spin-density waves, charge and nematic order, and high-temperature superconductivity \cite{Keimer2015} (SC), emanating from the combination of complex interactions that govern a simple two-dimensional (2D) chemical structure --- the CuO$_2$ planes.
These interactions within the CuO$_2$ planes are largely influenced by the underlying characteristics of the anisotropic, planar Cu 3$d_{x^2-y^2}$ orbitals, which dominate the density of states near the Fermi surface due to a sizeable energy splitting of the $e_g$ orbitals \cite{Tokura2000}.
The 2D character of the surviving $d_{x^2-y^2}$ orbital is observable in both transport and scattering measurements, for example, evidenced by its anisotropic electrical and thermal conductances\cite{Sato_1995} and overwhelmingly broad scattering peaks along $L$ in reciprocal space\cite{Chang_2016}, respectively. Understanding how the dimensionality of the orbital degrees of freedom affect the stability and interplay of these phases could reveal important information about the mechanism of superconductivity with broader applications for modifying the characteristics of correlated oxides and layered materials via orbital engineering.

\begin{figure*}[ht]
    {\includegraphics[width=\textwidth]{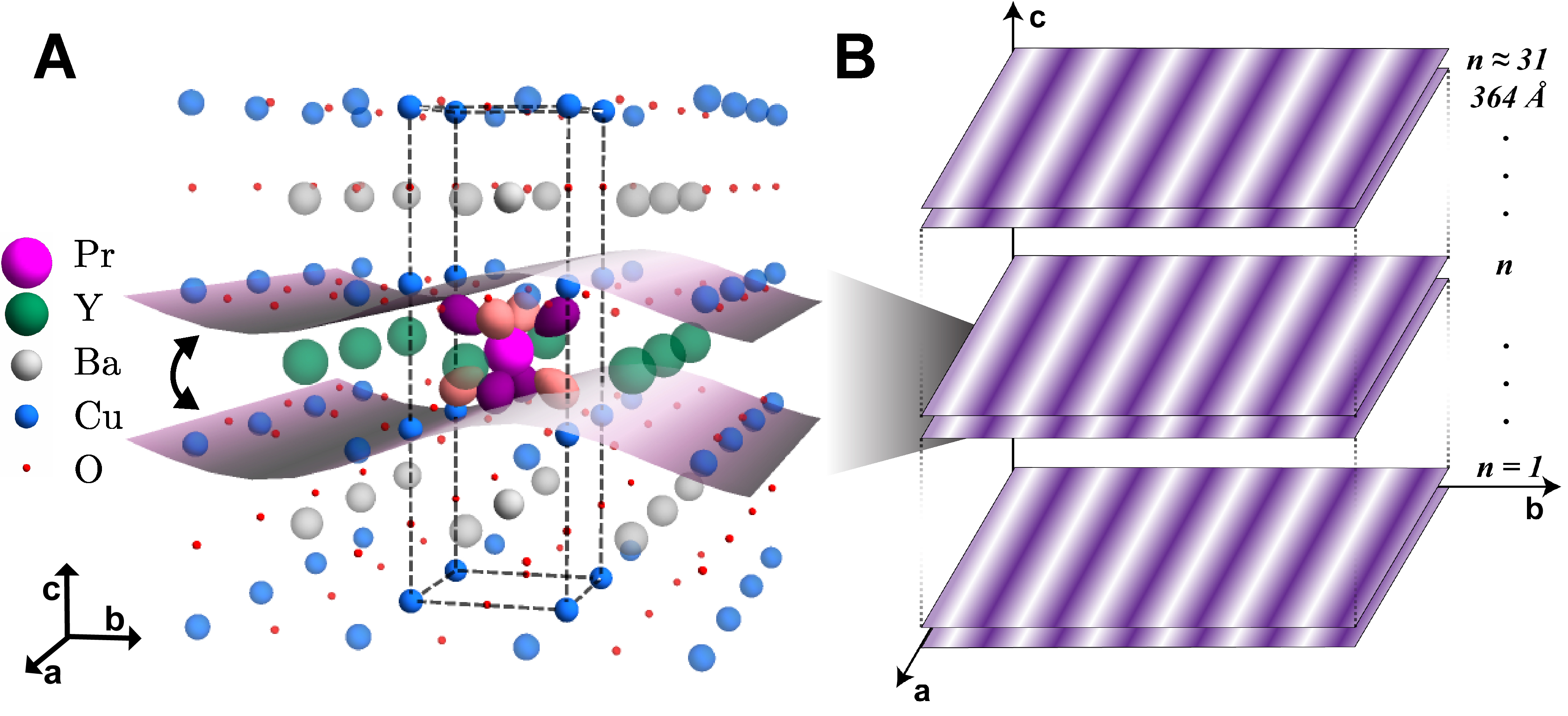}
  	\caption{\textbf{Stabilizing three dimensional charge order.} (A) The extended unit cell (dashed box) of  \prybcosev~illustrating the CO coupling between adjacent CuO$_2$ planes that arises when introducing Pr at the Y sites. The Pr 4$f_{z(x^2-y^2)}$ orbital is shown which hybridizes with the 2$p_\pi$ states of the planar oxygens. (B) A schematic depiction of the 3D CO out-of-plane correlation length in \prybcosev~which spans approximately 31 sets of CuO$_2$ planes ($\sim$364 \AA). }
  	\label{fig:fig1}} 
\end{figure*}

Among the most important cases of this interplay is the phenomenon of charge order (CO), a phase that is closely interconnected with SC\cite{Frano_2020,Comin_review, tranquada_evidence_1995, Hoffman_2002, Howald_2003, Vershinin_2004, Abbamonte_2005, Ghiringhelli_CDW_2012, Chang_CDW_2012, Comin_CDW_2014, daSilvaNeto_CDW_2014, Tabis_2014, Santi_2014, daSilvaNeto_Science_2015}. Incommensurate CO exists in all superconducting cuprates as a 2D electronic phenomenon hosted in the CuO$_2$ planes, reflecting the weak interplanar coupling of the planar Cu 3$d_{x^2-y^2}$ orbitals. In diffraction experiments, the 2D character is evidenced by a reciprocal space `rod' that is broad along the out-of-plane direction (Miller index $L$), and maximized at half-integer values of $L$ due to a weak, but out-of-phase, coupling between adjacent planes~\cite{forgan_microscopic_2015}. The strength of this interplanar coupling can be further quantified by extracting the correlation lengths from the widths of the scattered CO peaks along $L$. In \ybco~(YBCO), the highest reported out-of-plane correlation length (10 \AA) is nearly an order of magnitude smaller than the highest reported in-plane correlation length (95 \AA)\cite{Frano_2020}, highlighting the 2D nature of the CO phase. It is not clear whether disorder~\cite{Wu_2011, Leyraud2007, Wu_2015} or the low dimensionality of the underlying Cu 3$d_{x^2-y^2}$ orbitals intrinsically limits the out-of-plane correlation length, or if the CO could, in principle, develop into a truly long-range order, as suggested by recent experiments~\cite{Wu_2011, Leyraud2007, Wu_2015}.

It has since been observed that the application of certain perturbations --- high magnetic fields\cite{Chang_2016, Gerber_2015, Jang2016}, epitaxial strain\cite{Bluschke2018}, or uniaxial stress\cite{Kim1040,Kim2021_PRL} --- can induce a CO phase with three-dimensional (3D) coherence. Upon the application of these external influences, a second CO peak emerges, this time centered at integer $L$-values, evidencing an out-of-plane coupling that locks the phase of adjacent CuO$_2$ planes. The 3D CO peaks have significantly increased out-of-plane correlation lengths, achieving up to 55 \AA\cite{Jang2016}, 61 \AA\cite{Bluschke2018}, and 94 \AA\cite{Kim1040}, respectively. All of these 3D CO correlation lengths are still considerably shorter than the typical crystalline $c$-axis correlation lengths found in this compound. Furthermore, the 2D rod centered at half-integer $L$-values gets \textit{enhanced} upon applying the external influences, showing a persistent coexistence of the 3D and 2D COs. While it is easy to discern the 2D nature of the unperturbed CO upon consideration of the underlying planar Cu 3$d_{x^2-y^2}$ orbitals, the mechanisms by which these external perturbations are able to add a 3D CO peak remain unclear. Moreover, application of these perturbations in-situ presents complicated technical challenges that preclude many experimental techniques altogether, making it difficult to systematically investigate how the dimensionality of the CO can be tuned and obscuring its connection with SC. Taking an orthogonal route, we hypothesized that 3D CO could instead be stabilized by virtue of tuning the underlying orbital character via hybridization to more directly enhance the out-of-plane coupling between adjacent CuO$_2$ plane layers.

\begin{figure*}[ht]
    {\includegraphics[width=\textwidth]{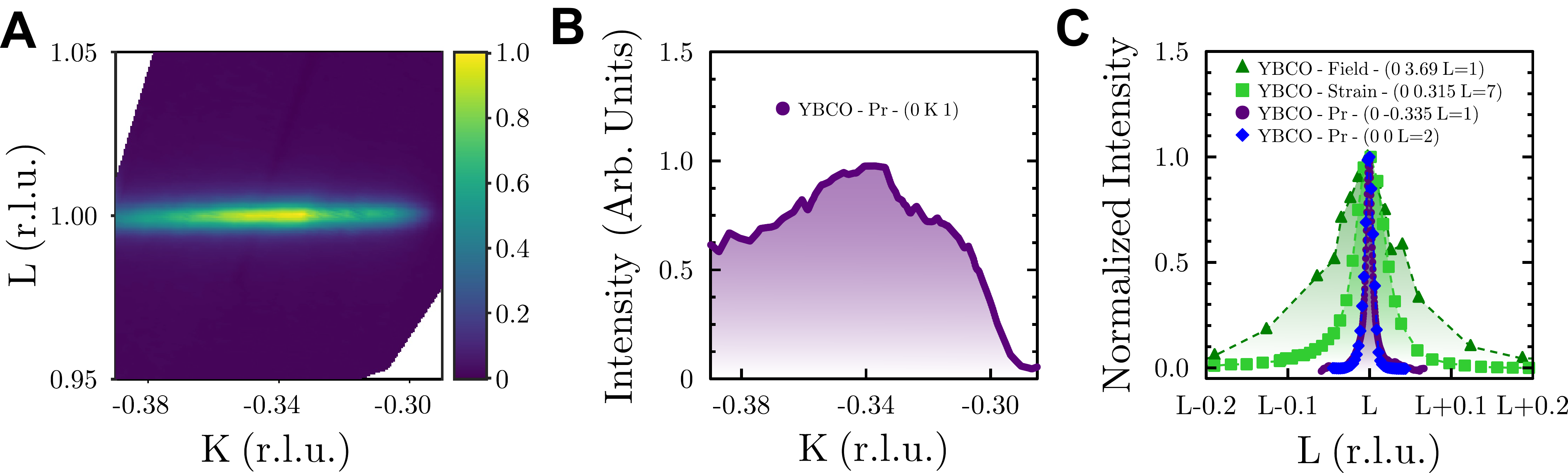}
  	\caption{\textbf{The reciprocal space structure of the 3D CO at T=50\,K.} (A) A $KL$ reciprocal space map collected at the Pr $M_5$ edge shows a diffraction feature centered at (0 -0.335 1) reciprocal lattice units (rlu). (B) A cut along $K$ at $L$=1 shows the peak is centered at an almost-commensurate value of $K$=-0.335. (C) A comparison of the out-of-plane $L$-widths of the known 3D CO peaks stabilized by magnetic field\cite{Chang_2016}, uniaxial strain\cite{Kim1040}, and the present work. For comparison, an $L$-cut of the structural (002) peak from \prybcosev~is shown. }
  	\label{fig:fig2}} 
\end{figure*}

Here we show that, by substituting Pr on the Y sites in \prybcosev ~(Pr-YBCO) (Figure \ref{fig:fig1}A), a highly correlated 3D CO state can be stabilized with an out-of-plane correlation length of $\sim$364 \AA ~(Figure \ref{fig:fig1}B), a number that is bound by the crystalline correlation length. This material was chosen because of how the hybridization between the Pr 4$f$ orbitals and the planar CuO$_2$ states\cite{Maple1994} yields an electronically relevant, hybridized orbital\cite{Yu_1998} with spatial extension in three dimensions, in stark contrast to the planar Cu 3$d_{x^2-y^2}$ orbitals that dominate the physics of the parent compound. Unlike substitution by other rare-earth elements, such as Dy, which do not significantly alter the parent YBCO phase diagram\cite{Betto_DyBCO}, increasing Pr-substitution in the \prybcosev~system continually reduces the superconducting $T_c$, yielding a pseudogap regime\cite{Sandu2004, Maple1994,Lobo_2001} and eventually an antiferromagnetic insulating phase\cite{Soderholm1987, Liang1987, Dalichaouch1988,Neumeier1989, Peng1989, Maple1994, radousky_1992}. Various results suggest that localized Pr 4$f$ states are appreciably hybridized with the valence band states associated with the conducting CuO$_2$ planes, specifically the oxygen 2$p$ level~\cite{Neumeier1988, Maple1992}. We present density-functional calculations showing that, through this hybridization which is unique to Pr, the CO on adjacent CuO$_2$ planes can couple to yield a stable 3D CO phase. Altogether, our results constitute the first detection of a fully stabilized, long-range 3D CO that competes with SC, achieved by intrinsically engineering the orbital character of the electronic structure.

We used resonant soft x-ray scattering (RSXS) at the Cu $L_3$ and Pr $M_5$ edges to investigate the CO properties in a \prybcosev~sample with $x\approx$0.3 and a superconducting $T_c$=50\,K; a concentration value chosen because it features pseudogap behavior, as measured by various probes~\cite{Sandu2004, Maple1994,Lobo_2001}, and because it yields a $T_c$ similar to underdoped YBa$_2$Cu$_3$O$_{\sim6.67}$, a doping level where the CO phase is maximal. A reciprocal space map of the $KL$-plane in reciprocal lattice units (rlu), measured at the Pr $M_5$ edge (930.3 eV) at 50\,K, is shown in Figure \ref{fig:fig2}A. In eminent contrast to all other reports of 3D CO, no scattered intensity was detected in the vicinity of $L\approx$1.5, indicating the apparent absence of 2D CO (more details in the Supplementary Information). This represents the first unique aspect of our work: to within the limits of our instrumental resolution, we only observe a peak at $L$=1, suggesting an effective isolation of the CO phase with an out-of-plane coupling.

Further inspection of the 3D CO signal displays a reciprocal space structure that is broad along $K$ but narrow along $L$. The broad shape of the peak along $K$ at $L$=1, shown in Figure \ref{fig:fig2}B, is consistent with the broad feature observed in many previous RSXS measurements of CO in cuprates~\cite{Ghiringhelli_CDW_2012, Comin_CDW_2014, daSilvaNeto_CDW_2014} that has been attributed to a fluctuating component in YBCO~\cite{Arpaia_Science}, suggesting that the actual static contribution may be narrower than it appears. Interestingly, the peak is centered at an almost-commensurate in-plane momentum value $K$=-0.335 rlu. Due to not having detwinned samples (Supplementary Information), we cannot determine whether the 3D CO peak is uniaxial.
Another important feature of our discovery is shown in Figure \ref{fig:fig2}C, which compares reciprocal space cuts along $L$ close to integer values with $K$ centered at the in-plane CO wavevector. The broadest peak (dark green triangles) displays the data reported for 3D CO induced by high magnetic field\cite{Chang_2016}. The next broadest peak (lime green squares) displays the data measured under the application of 1.0\% uniaxial strain\cite{Kim1040}, which has yielded 3D CO with the previously highest reported out-of-plane correlation length. Finally, the Pr-YBCO 3D CO peak (purple circles) is shown with the (002) structural reflection (blue diamonds) overlaid,  which was measured under very similar scattering conditions. The widths of these two peaks are within the margin of error of each other. Thus, in this Pr-YBCO system, the 3D CO peak has a width that is resolution-limited by the width of the crystallographic Bragg peaks. The resulting lower bound on the out-of-plane correlation length is 364 {\AA}.

\begin{figure*}[ht]
    {\includegraphics[width=\textwidth]{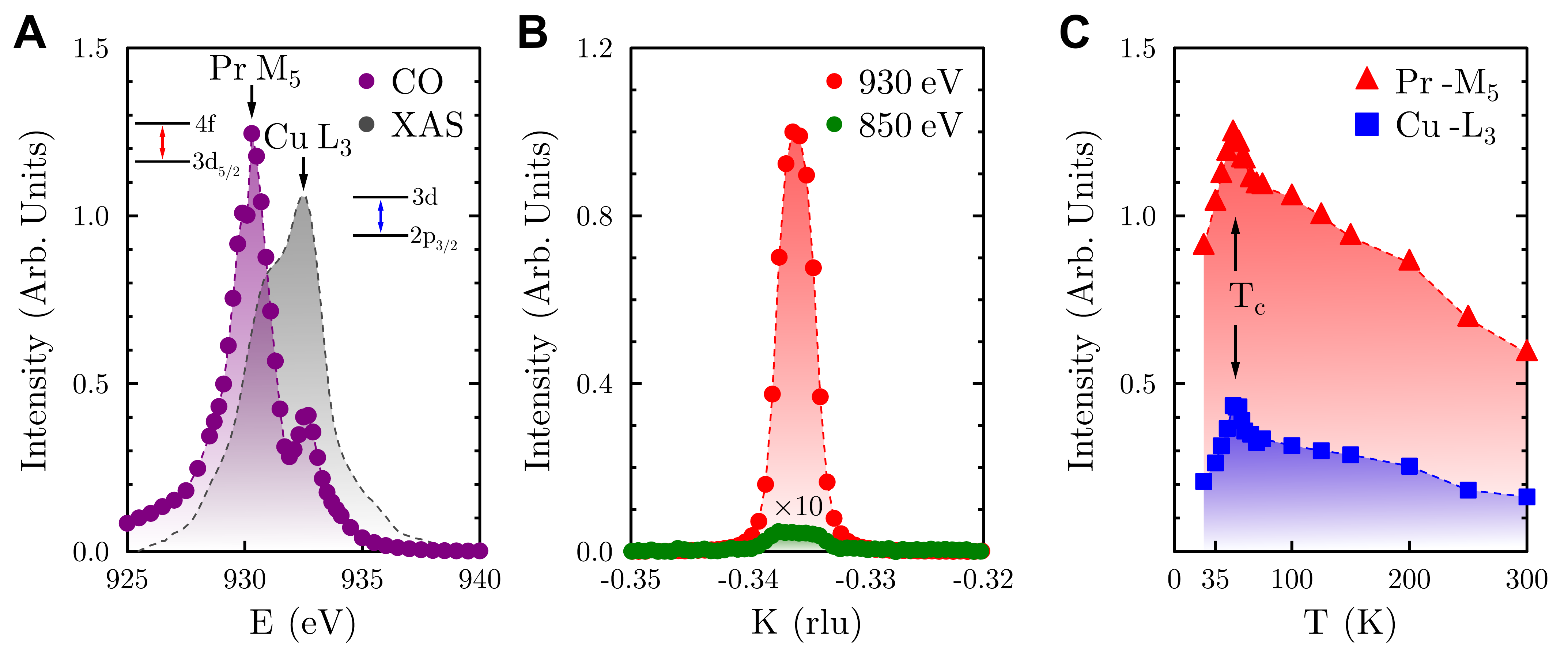}
  	\caption{\textbf{The nature of the 3D CO.} (A) The XAS (gray) shows the Cu $L_3$ edge with a shoulder at lower energies corresponding to the Pr $M_5$ edge. The two dipole-allowed transitions are labeled as insets. The purple data show the energy dependence of the scattered intensity of the 3D CO at T=50\,K. (B) Rocking curve scans of the 3D CO peak taken on resonance (930\,eV) and well below the resonance (850\,eV), the latter is enhanced 10 times to show a weaker, but detectable, off-resonant intensity. (C) The temperature dependencies at the Cu and Pr resonances, showing a detectable peak at room temperature and a drop below the SC transition $T_c$. }
  	\label{fig:fig3}} 
\end{figure*}

The energy dependence of the scattered intensity at (0 -0.335 1) is shown in Figure \ref{fig:fig3}A, overlaid with the corresponding x-ray absorption spectrum (XAS) measured with the electric field of the x-rays parallel to the bond directions in the CuO$_2$ planes. The XAS reveals two resonances that correspond to the Pr $M_5$ ($\sim$930\,eV) and the Cu $L_3$ ($\sim$933\,eV) edges; the 3D CO scattered intensity also exhibits features corresponding to these two resonances. Unlike in YBCO films with 3D CO\cite{Bluschke2018}, we do not observe a significant shift in spectral weight to higher energy that would indicate CO coupling through the CuO chains. 
Notably, the CO peak resonates more strongly at the Pr edge than at the Cu edge; while it is not straightforward to draw quantitative conclusions from this result, the intense scattering signal nonetheless confirms involvement by the Pr ions in the 3D CO mechanism. Furthermore, we observe in Figure \ref{fig:fig3}B that the 3D CO peak can still be detected at energies far below the resonance (850\,eV) albeit much more weakly, which is in contrast to all other cuprates where the CO peaks studied by RSXS lack sufficient scattering strength to be observed off-resonance. This indicates a sizeable lattice distortion unseen in other cuprate systems, highlighting that in Pr-YBCO the 3D CO becomes more structurally stable than previously reported.

\begin{figure*}[ht]
    {\includegraphics[width=.99\textwidth]{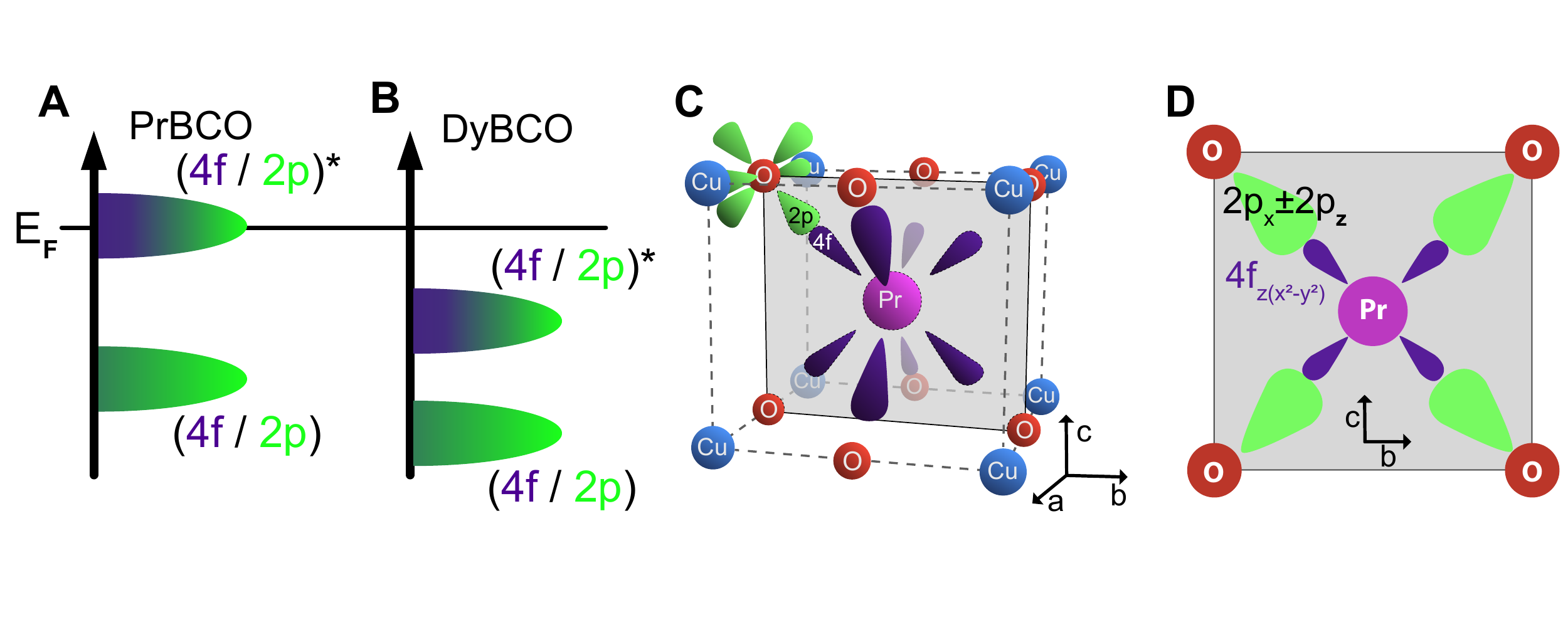}
  	\caption{\textbf{DFT-calculated band structures showing hybridization.} (A) A schematic representation of the orbital character of the electronic levels near the Fermi energy ($E_F$) in \prbcosix, showing a mixed 4$f$ (purple) and 2$p$ (green) antibonding band crossing $E_F$. (B) The equivalent schematic for \dybcosix. (C) A schematic depicting the 4$f$ (purple) and 2$p$ (green) orbitals within the crystal structure of \prbco~in three dimensions. (D) The hybridization that occurs between the 4$f_{z(x^2-y^2)}$ (purple) and a linear combination of 2$p_x$ and 2$p_z$ orbitals (green) oriented towards each other, shown within the plane represented in panel (C).}
  	\label{fig:dft}} 
\end{figure*}

The interplay of the 3D CO with superconductivity can be investigated by measuring the temperature dependence of the former. In Figure \ref{fig:fig3}C, we plot the scattered intensity at (0 -0.335 1) at both resonance energies as a function of temperature. While the overall scattering intensity is higher at the Pr $M_5$ energy than the Cu $L_3$, which is consistent with the measured energy dependence, it is notable that the 3D CO scattering signal is still detectable at room temperature for both energies. Upon cooling from T=300\,K, the temperature dependence at both energies maintain roughly equivalent slopes until within the vicinity of $T_c$=50 K. Cooling below $T_c$ produces a cusp-like maximum, indicating a competition between SC and the isolated 3D CO phase. This is in contrast to the 3D CO induced by very high magnetic fields, where any competition between 3D CO and SC is obscured by the very presence of the magnetic field which, while necessary to induce 3D coherence, comes with the unavoidable expense of greatly suppressing the SC phase.

Having established experimentally that 3D CO can be stabilized with long out-of-plane correlation length, we turn to discuss the possible origin of the $c$-axis coupling in the Pr-YBCO system. It is already well known that, unlike any other rare earth, Pr-substitution uniquely suppresses SC in \ybco~by localizing holes via orbital hybridization~\cite{Neumeier1988, Maple1992,Fehrenbacher1993,Liechtenstein1995}. To this end, we performed density- functional theory plus Hubbard U (DFT+U) calculations for both \prbcosix~and \dybcosix~structures~\cite{dft_struct} to understand the role of this hybridization within the context of 3D CO and its competition with SC (details in the Supplementary Information). Figure~\ref{fig:dft}A schematically depicts the orbital character of the electronic states near the Fermi level ($E_F$) in \prbcosix. In addition to the characteristic $pd\sigma$ bands of the CuO$_2$ planes which host all the 2D electronic phenomena, another band crosses the Fermi level. This results from the antibonding coupling between the Pr 4$f_{z(x^2-y^2)}$ state and its nearest-neighbor O 2$p_\pi$ states in adjacent CuO$_2$ planes~\cite{Fehrenbacher1993,Liechtenstein1995} (Figure~\ref{fig:dft}C-D). We speculate that this orbital coupling with an out-of-plane component locks together the phase of the CO on adjacent CuO$_2$ planes, resulting in a diffraction peak at $L$=1~\cite{forgan_microscopic_2015}. For later rare-earth elements with lower 4$f$ energy, the 4$f_{z(x^2-y^2)}$--2$p_\pi$ antibonding band is expected to be lowered and removed from the Fermi level. Figure~\ref{fig:dft}B depicts the calculated electronic level structure for \dybcosix, where the top of this band is deep below the Fermi level at around $-0.8$ eV. This leaves the charge carriers in the two adjacent CuO$_2$ planes essentially decoupled. This unique aspect of Pr makes it the appropriate rare-earth to substitute into Y to stabilize and isolate 3D CO, which according to our data occurs concomitantly with a lattice distortion. The exact structural mechanism of stabilization, e.g., phonons, is a subject of future research.

Our discovery of a fully stable 3D CO without a 2D signal has important implications to our understanding of CO and its interplay with SC. First, we confirm that a fully coherent, isolated 3D CO can be stabilized despite the intrinsic disorder inevitably present in cuprates. We note here that our Pr-substituted samples are expected to host at least as much structural and chemical disorder than in pristine YBCO, if not more so, due to the additional defect channel. Second, we conﬁrm that a stable 3D CO still coexists and competes with SC, implying that the system's ground state can comprise two long-range, static, coexisting orders. Third, the stable 3D CO manifesting at almost-commensurate momentum values is consistent with the speculation that, when fully crystallized, the CO will shift and lock into a commensurate structure\cite{forgan_microscopic_2015}. Fourth, since the 3D-coupling does not rely on the CuO chains that are unique to YBCO, perhaps other forms of hybridization can be used to stabilize 3D CO in other cuprate families, which has not yet been observed. Finally, we show that controlling the orbital content of the Fermi surface by assigning it a 4$f$ character with an out-of-plane component can yield a sizable impact on the electronic ordering tendencies of the CuO$_2$ plane. One can use it as a tuning knob to study the validity of single-band models used to describe 2D systems like the cuprates or intercalated graphitic systems ~\cite{PhysRevLett.102.107007,Yang2019,Chen2019,Cao2018_correlatedMAG, Cao2018_SC_MAG}.

In summary,  we have shown how utilizing the hybridization between the 4$f$ states of Pr and  planar CuO$_2$ orbitals to tune the underlying orbital character can significantly enhance the out-of-plane coupling, phase-locking the CO across adjacent planes and rendering a stable CO phase that is fully correlated along the out-of-plane direction without the 2D version. The $c$-axis correlation length has a lower bound matching that of the crystal itself, showing that Pr-substitution is the most efficient way of stabilizing 3D CO compared to using external perturbations, like magnetic fields and strain, and uniquely does not suffer from experimental complications arising from in-situ application. Furthermore, through resonant spectroscopy, we attribute the formation of 3D coupling to the role of the Pr ions located between CuO$_2$ planes. To understand the mechanism of this out-of-plane coupling, we turned to DFT+U calculations that show a hybridized 4$f$-2$p$ band crossing the Fermi level, a feature that is unique to Pr-doped YBCO. Since our system does not rely on external perturbations, other techniques can be employed to investigate this material and shed light on the connection between CO and SC. Moreover,  this demonstrates how the influence of underlying orbital character on an electronic phase can be tuned via orbital hybridization, which can be generalized to other correlated transition metal oxides and layered systems.

\noindent \textbf{Acknowledgements}\\
\small
We thank Bernhard Keimer, Davide Betto, Fabio Boschini, George Sawatzky, Martin Bluschke, Matteo Minola, Mingu Kang, Riccardo Comin for fruitful discussions.
Research at UC San Diego (M.B.M., K.S., C.M.M.) was supported by the US Department of Energy, Office of Basic Energy Sciences, Division of Materials Sciences and Engineering, under Grant No. DEFG02-04-ER46105 (single crystal growth) and US National Science Foundation under Grant No. DMR-1810310 (materials characterization).
Resonant soft x-ray experiments were carried out at the SSRL (beamline 13–3), SLAC National Accelerator Laboratory. This study at the SSRL/SLAC is supported by the U.S. Department of Energy, Office of Science, Office of Basic Energy Sciences under contract no. DE- AC02-76SF00515. This research also used resources of the Advanced Photon Source (29ID), a U.S. Department of Energy (DOE) Office of Science User Facility operated for the DOE Office of Science by Argonne National Laboratory under Contract No. DE-AC02-06CH11357.
Y.L. acknowledges support by Deutsche Forschungsgemeinschaft (DFG) under Germany’s Excellence Strategy EXC2181/1-390900948 (the Heidelberg STRUCTURES Excellence Cluster).
Y.H. acknowledges previous support from the Miller Institute for Basic Research in Sciences.
S.B-C acknowledges support from the MINECO of Spain through the project PGC2018-101334-AC22.
E.H.d.S.N. acknowledges previous support from UC Davis startup funds, as well as current support from the Alfred P. Sloan Fellowship in Physics.

\noindent \textbf{Author contributions}\\
\small
A.F. and M.B.M. conceived and led the project. The RSXS experiments were performed by A.R., B.G., H. H., J.-S.L., F.R., T.J.B., M. W., Y.H., S.B.-C., and E.H.d.S.N. Single crystals were grown and characterized by K.S., C.M.M., and M.B.M. The data analysis was carried out by A.R., B.G., H. H., and J.-S.L. DFT+U calculations performed by Y.L. The manuscript was written by A.F., A.R., B.G., Y.H., E.H.d.S.N, and S.B.C. with input from all the co-authors. 

\noindent \textbf{Competing interests:}\\
The authors have no competing interests.

\noindent \textbf{Data and materials availability:}\\
All raw data are available at this \href{https://doi.org/10.7910/DVN/2BIWWI}{link}


%

\end{document}